\documentclass[aps,pra,twocolumn,nopreprintnumbers,noeprint,superscriptaddress]{revtex4-2}
\usepackage{graphicx}
\usepackage{dcolumn}
\usepackage{bm}
\usepackage{amsmath}
\usepackage{amsfonts}
\usepackage{psfrag}
\usepackage{graphicx}
\usepackage{color} 
\usepackage{footmisc}
\usepackage{mathtools}
\usepackage{amssymb}
\usepackage[normalem]{ulem}
\usepackage[english]{babel}

\usepackage{hyperref}

\begin{document}
 
\title{Perspective on real-space nanophotonic field manipulation using non-perturbative light-matter coupling}

\author{Erika Cortese}
\email{E-mail: e.cortese@soton.ac.uk}
\affiliation{School of Physics and Astronomy, University of Southampton, Southampton, SO17 1BJ, United Kingdom}

\author{Joshua Mornhinweg} 
\affiliation{Department of Physics, University of Regensburg, 93040 Regensburg, Germany} 

\author{Rupert Huber} 
\affiliation{Department of Physics, University of Regensburg, 93040 Regensburg, Germany} 

\author{Christoph Lange}
\affiliation{Department of Physics, TU Dortmund University, 44227 Dortmund, Germany} 

\author{Simone \surname{De Liberato}}
\email{E-mail: s.de-liberato@soton.ac.uk}
\affiliation{School of Physics and Astronomy, University of Southampton, Southampton, SO17 1BJ, United Kingdom}

\begin{abstract}
The achievement of large values of the light-matter coupling in nanoengineered photonic structures can lead to multiple photonic resonances contributing to the final properties of the same hybrid polariton mode.
We develop a general theory describing multi-mode light-matter coupling in systems of reduced dimensionality and we explore their novel phenomenology, validating our theory’s predictions against numerical electromagnetic simulations. 
On the one hand, we characterise the spectral features linked with the multi-mode nature of the polaritons. On the other hand, we show how the interference between different photonic resonances can modify the real-space shape of the electromagnetic field associated with each polariton mode.
We argue that the possibility of engineering nanophotonic resonators to maximise the multi-mode mixing, and to alter the polariton modes via applied external fields, could allow for the dynamical real-space tailoring of subwavelength electromagnetic fields.
\end{abstract}

\maketitle

\section{Introduction}
Confining light below the Abbe diffraction limit \cite{ballariniPolaritonicsMicrocavitiesSubwavelength2019} by storing a part of the electromagnetic energy in the kinetic energy of electric charges \cite{Khurgin2018} opened the door to a number of groundbreaking real-world applications which has contributed to the great success of the field of nanophotonics.
In a nanophotonic device, the high energy density of the electromagnetic field makes it relatively easy to couple with different kinds of localised material excitations and reach the strong light-matter coupling regime, originally achieved in cavity quantum electrodynamics (CQED) atomic systems \cite{haroche_exploring_2006}.
In such a regime, light and matter degrees of freedom hybridise, leading to novel, polaritonic excitations of mixed light-matter character \cite{kavokinMicrocavities2017,Basov2021}.
 
Standard theoretical models used to describe strong coupling consider a single optically active matter transition coupled to a single photonic mode. 
Although some care has to be used when performing calculations on such a reduced Hilbert space \cite{debernardisBreakdownGaugeInvariance2018,sanchezmunozResolutionSuperluminalSignalling2018,distefanoResolutionGaugeAmbiguities2019,stokesGaugeAmbiguitiesImply2019}, this single-mode approximation has enabled modelling of a wide range of CQED systems with remarkable easiness and generality. However, the requirement is that the energy spacing between the considered resonances and the neglected ones is much larger than the strength of light-matter coupling, thus permitting to integrate out excited modes with negligible populations.

 However, the ongoing race for record coupling strengths \cite{kockum2019ultrastrong,SolanoRMP} has led to situations in which higher-energy electronic states cannot be neglected, requiring a model which considers the coupling of multiple matter excitations to the same photonic mode. We refer to this regime as the very-strong coupling (VSC) regime, first predicted by Khurgin in 2001 \cite{khurginExcitonicRadiusCavity2001}.
 The hybridization of multiple excited matter states has an important consequence: the matter component of the polariton, represented itself by a linear superposition of different bare matter wavefunctions, has a wavefunction different from each of the bare states \cite{Khurgin2019}. Following a 2013 proposal \cite{Zhang2013}, such an effect was observed for the first time in 2017 \cite{brodbeckExperimentalVerificationVery2017}, as a modification of approximately $30\%$ of the Wannier exciton Bohr radius in GaAs microcavities. Larger numbers of matter states which can be hybridised by the coupling with the photonic field could correspond to a broader design space for the resulting electronic wavefunction. This idea led to the study of systems with a continuum of ionised excitations \cite{Cortese:19,Cortese:2022} and eventually to the discovery of novel bound excitons stabilised by the photonic interaction \cite{cortese_excitons_2021}, and to novel polaritonic loss channels \cite{rajabali2021polaritonic}.

\begin{figure}[htbp]
\begin{center}
\includegraphics[width=0.5\textwidth]{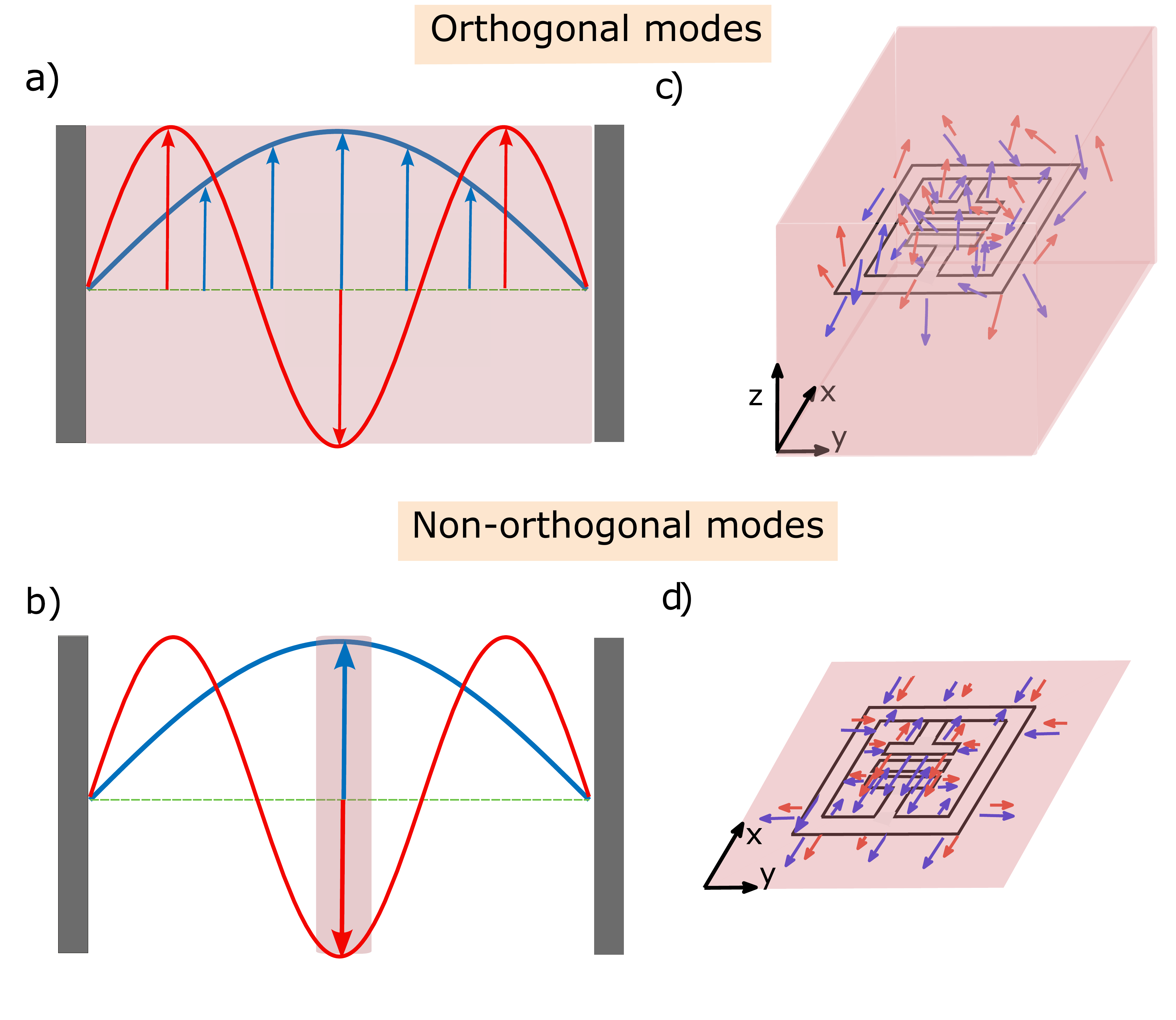}
\caption{\label{Fig1} 
 Sketch of how orthogonal resonator modes can become non-orthogonal when coupled over an active region occupying only part of the resonator volume. Shown are the case of a planar microcavity (a,b) and a split-ring resonator (c,d). Two electromagnetic modes are shown by red and blue arrows, and the
 active region, corresponding to the full three-dimensional volume (a,c) or a thin, quasi-two-dimensional surface (b,d) is shaded in light red.}
\end{center}
\end{figure}
 In this Article we theoretically investigate the possibility of both multi-mode electronic as well as multi-mode photonic hybridization, leading to a modification of the spatial electromagnetic profiles of the resulting polariton modes. Given the possibility of fast \cite{Anappara2005,schwartz2011} in-situ tuning of the light-matter interaction by optical and electrical means, subcycle multiwave mixing nonlinearities between different polariton states \cite{Mornhinweg2021} or even all-optical subcycle switching \cite{gunterSubcycleSwitchonUltrastrong2009,Halbhuber2020},
 such an approach could open the door to dynamical manipulation of subwavelength fields, with potential disruptive applications for, e.g., on-chip optical tweezers \cite{Yu2021}.  

Although to the best of our knowledge it was never explicitly discussed, the regime of photonic VSC has been already reached in various systems, as superconducting qubits coupled to microwave photons in a long transmission-line resonator \cite{sundaresan2015beyond,liu_quantum_2017}.
Moreover, it has been theoretically \cite{deliberatoLightMatterDecouplingDeep2014} and experimentally \cite{bayerTerahertzLightMatter2017} demonstrated in microcavities, where the coupling strength becomes larger than the bare excitation frequencies. In such a regime, the diamagnetic term of the Hamiltonian creates a dominant real-space repulsive interaction localised at the dipole position, which expels the electromagnetic field and may even lead to light-matter decoupling \cite{deliberatoLightMatterDecouplingDeep2014, bayerTerahertzLightMatter2017}. It has also been experimentally observed that, in plasmonic nanocavities, the greatly enhanced coupling between molecular excitons and gap plasmons causes a significant modification of the plasmonic modes profile \cite{chenModeModificationPlasmonic2017}.

Here we focus on Landau polaritons, where the giant electronic dipoles of cyclotron resonances (CRs) of two-dimensional electron gases (2DEGs) are coupled to strongly enhanced light fields of subwavelength THz resonators. After inital predictions in Ref. \cite{hagenmullerUltrastrongCouplingCavity2010}, multiple experimental realizations followed, some of which established world-records for the largest light-matter coupling ever achieved in any CQED system \cite{scalariUltrastrongCouplingCyclotron2012,Scalari2013,bayerTerahertzLightMatter2017,liVacuumBlochSiegert2018}. 

In the first part of the paper we will develop a general theory describing multi-mode light-matter coupling of CRs of a planar 2DEG which is Landau-quantised by a perpendicular static magnetic field, and multiple photonic resonances. Our approach highlights the main electronic and optical features observable for this multi-mode coupling. 
In the second part, we apply our formalism to structures based on planar plasmonic metasurfaces.
To this end, we perform numerical simulations which verify the predictions of our theory and demonstrate how multi-mode photonic hybridization can lead to a modification of the electromagnetic spatial profile of the polariton modes.

\section{Theory of multi-mode light-matter coupling}
\begin{figure}[htbp]
\begin{center}
\includegraphics[width=0.45\textwidth]{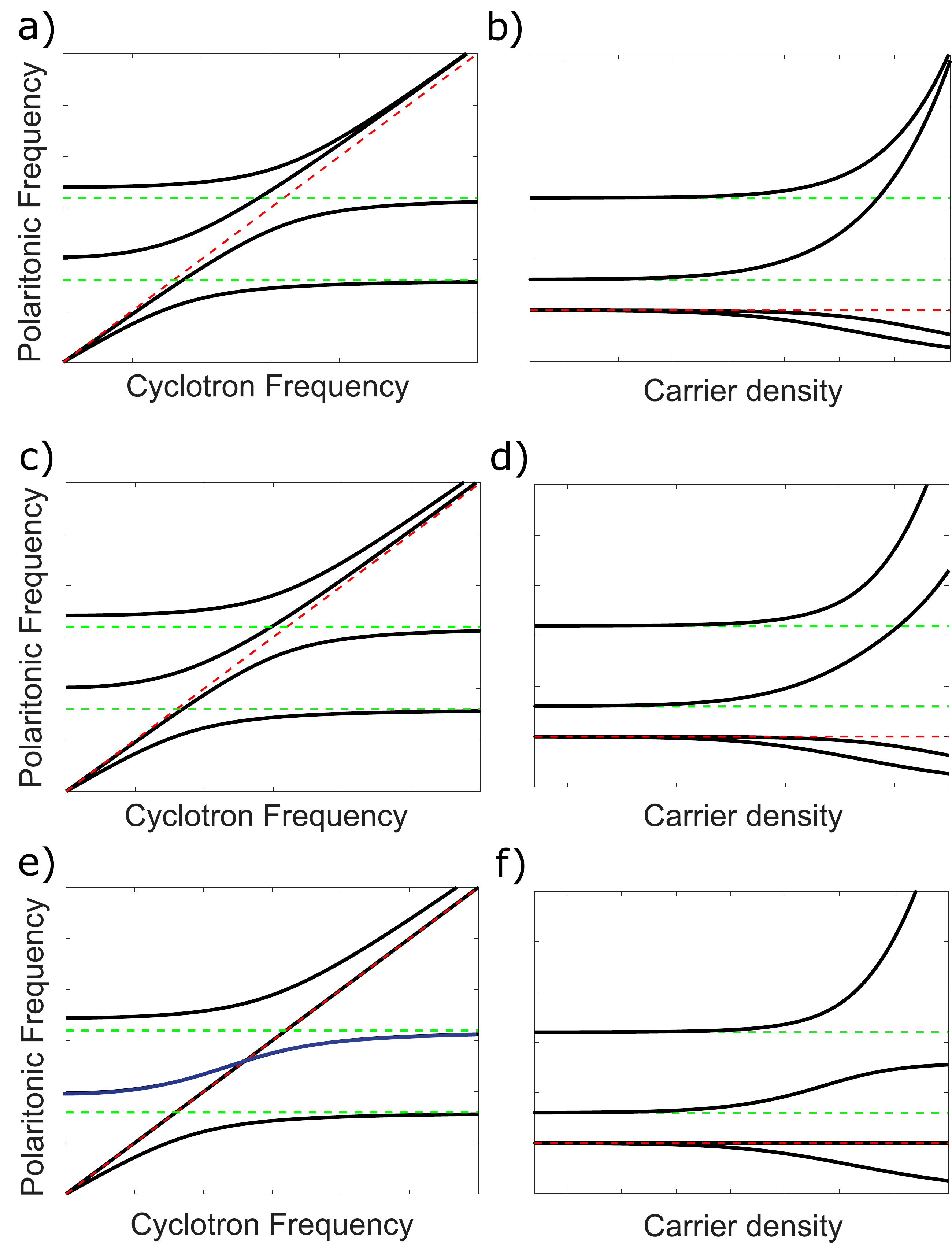}
\caption{\label{Fig2} 
 Polaritonic eigenmodes arising from the diagonalization of Eq.~\ref{finalH} with $M=2$ photonic resonances (green dashed lines) coupled to 2DEG with cyclotron resonance $\omega_c$ (red dashed line). 
 The three rows correspond to the case of zero overlap between the two photonic modes ($\eta_{2,1}=0$ a,b),
 medium overlap ($\eta_{2,1}=0.5$ c,d), or perfect overlap ($\eta_{2,1}=1$ e,f).
 Panels on the left column (a,c,e) are shown as a function of the cyclotron frequency $\omega_c$ while those on the right one as a function of the electron density $N_{\textrm{2DEG}}$. Panel (e) displays a \emph{S-shaped} polariton curve (blue solid line) due to a perfect overlap.}
\end{center}
\end{figure}
In this section we develop a theory for the light-matter coupling between $M$ photonic resonator modes and the CRs of a 2DEG with a charge carrier density $N_{\textrm{2DEG}}$ and an effective mass $m^*$. Following Kohn's theorem \cite{kohn1961cyclotron}, we neglect Coulomb interactions between the electrons, which manifest in the nonlinear susceptibility of strongly driven Landau electron systems \cite{MaagBeyondKohn2016}, but have no role in the determination of the optical resonance. Moreover, while our theory focuses on a single quantum well (QW) hosting the entire electron distribution, it is equally applicable to densely packed multi-QW structures as usually employed in experiments, where the intensity of the electromagnetic field doesn't vary significantly within the thickness of the multi-QW stack.
Following the elegant theory from Ref. \cite{liVacuumBlochSiegert2018} we can write the Hamiltonian of our system as
\begin{eqnarray} \label{H}
\hat{H}&=& \hat{H}_{\text{cav}}+\sum_{j=1}^N \hbar\omega_c\hat{c}_j^{\dagger}\hat{c}_j
+\frac{e^2}{m^*}\sum_{j=1}^N \hat{A}_-(\mathbf{r}_j)\hat{A}_+(\mathbf{r}_j)\nonumber\\&+& i\sqrt{\frac{\hbar\omega_c e^2}{m^*}}\sum_{j=1}^N\left[ \hat{c}_j^{\dagger}\hat{A}_+(\mathbf{r}_j)-
\hat{c}_j\hat{A}_-(\mathbf{r}_j)\right],
\end{eqnarray}
where $\hat{c}_j$ is the bosonic lowering operator for the electrons, leading to a transition from the $j$th to the $(j-1)$th Landau level with a transition energy $\hbar\omega_c$, and $\hat{H}_{\text{cav}}$ is the Hamiltonian describing the bare electromagnetic field in the resonator. 
In case of high electron density and strong in-plane confinement of both the 2DEG and the electromagnetic field, plasmonic modes hosted by the system can be non-negligible and lead to the formation of magneto-plasmon modes with a renormalized frequency of $\tilde{\omega}_c=\sqrt{\omega_c^2+\omega_P^2}$, where $\omega_P$ is the 2D plasmon frequency for the 2DEG \cite{rajabali2021polaritonic}. However, a correct estimation of $\omega_P$ not only takes into account the in-plane confinement of the 2DEG, but also includes the screening of the metallic resonator in proximity of the electrons, leading to a reduction of the plasmon energy \cite{popovResonantExcitationPlasma2005, bialekPlasmonDispersionsHigh2014, paravicini-baglianiGateMagneticField2017}.
For our structures, this effect strongly limits the extent of renormalization such that we disregard plasmon effects. 

In Eq. \ref{H} we introduced the non-Hermitian vector potentials written in terms of the in-plane component of the vector potential $\hat{\mathbf{A}}(\mathbf{r})$ as
\begin{eqnarray}
\hat{A}_{\pm}(\mathbf{r})&=&\frac{\hat{A}_{x}(\mathbf{r})\mp i\hat{A}_{y}(\mathbf{r})}{\sqrt{2}}.
\end{eqnarray}

The full vector potential can be expressed as a sum of photonic modes obtained by solving Maxwell's equations for the bare cavity, with dimensionless spatial field profiles $\bf{f}_\nu(\textbf{r})$, frequencies 
$\omega_\nu$, and second-quantized bosonic annihilation operators $\hat{a}_{\nu}$ as
\begin{eqnarray} \label{PotentialAfull}
  \hat{\mathbf{A}}(\textbf{r})&=&\sum_{\nu} \sqrt{\frac{\hbar}{2 \epsilon_0 \epsilon_r(\mathbf{r}) \omega_{\nu} \mathbb{V}_\nu}} \mathbf{f}_{\nu} (\textbf{r})\left (  \hat{a}_{\nu}^\dagger +  \hat{a}_{\nu} \right).
\end{eqnarray}
Here, the classical modes are orthogonal over the full integration domain $\mathbb{V}$  \cite{Gubbin2016}
  \begin{eqnarray} \label{V}
  \int_\mathbb{V}  \bf{f}^*_\nu(\mathbf{r}) \bf{f}_\mu(\mathbf{r})d\mathbf{r} =  \mathbb{V}_\nu\delta_{\nu,\mu},
  \end{eqnarray}
  with $\mathbb{V}_\nu$ the mode volume of the $\nu$th photon mode and $\epsilon_r(\mathbf{r})$ the background, non-resonant dielectric constant. The amplitudes of the non-Hermitian vector potentials then take the form
\begin{eqnarray} \label{PotentialA}
  \hat{A}_{-}(\textbf{r})&=&\sum_{\nu} \sqrt{\frac{\hbar}{2 \epsilon_0 \epsilon_r(\mathbf{r}) \omega_{\nu} \mathbb{V}_\nu}} f_{\nu} (\textbf{r})\left (  \hat{a}_{\nu}^\dagger +  \hat{a}_{\nu} \right),\\
  \hat{A}_{+}(\textbf{r})&=&\sum_{\nu} \sqrt{\frac{\hbar}{2 \epsilon_0 \epsilon_r(\mathbf{r}) \omega_{\nu} \mathbb{V}_\nu}} f^*_{\nu} (\textbf{r})\left  (   \hat{a}_{\nu}^\dagger +  \hat{a}_{\nu} \right), \nonumber
  \end{eqnarray}
  with 
   \begin{eqnarray}
   f_{\nu}(\mathbf{r})&=&\frac{f_{\nu,x}(\mathbf{r})+if_{\nu,y}(\mathbf{r})}{\sqrt{2}}.
  \end{eqnarray}

 Crucially, the orthogonality condition in Eq.~\ref{V} holds only if the integral is performed over the entire three-dimensional space, while the integral of two orthogonal modes over any sub-domain does not vanish in general. 
 This concept is illustrated in Fig. \ref{Fig1} for the model case of a planar microcavity (a,b) and for a split-ring resonator (c,d), integrated over either the full three-dimensional volume (a,c) or a thin, quasi-two-dimensional surface (b,d).
  In both cases, two orthogonal modes (red and blue arrows) become non-orthogonal when the integral is performed over a quasi-two-dimensional slice of the overall volume.
  This finding is relevant for our systems because the third term of Eq.~\ref{H}, the so-called diamagnetic term of the light-matter interaction Hamiltonian, contains generally non-vanishing expressions of the form
  \begin{eqnarray} \label{HD}
 \sum_{j=1}^N f^*_{\nu}(\mathbf{r}_j)
 f_{\mu}(\mathbf{r}_j)
 &=&
 N_{\textrm{2DEG}}\int_\mathbb{S} f^*_{\nu}(z,\mathbf{r}_{\|})
 f_{\mu}(z,\mathbf{r}_{\|}) d\mathbf{r}_{\|},\nonumber \\
  \end{eqnarray}
where $\mathbb{S}$ is the sample surface, $z$ is the out-of-plane position of the 2DEG and $\mathbf{r}_{\|}$ is the in-plane position.
 Placing a 2DEG at the center of the planar microcavity, or below the split-ring resonator, will thus result in an interaction of different photon modes which is mediated and modulated by the coupling to the electrons.
 
\begin{figure}[htbp]
\begin{center} 
\includegraphics[width=0.5\textwidth]{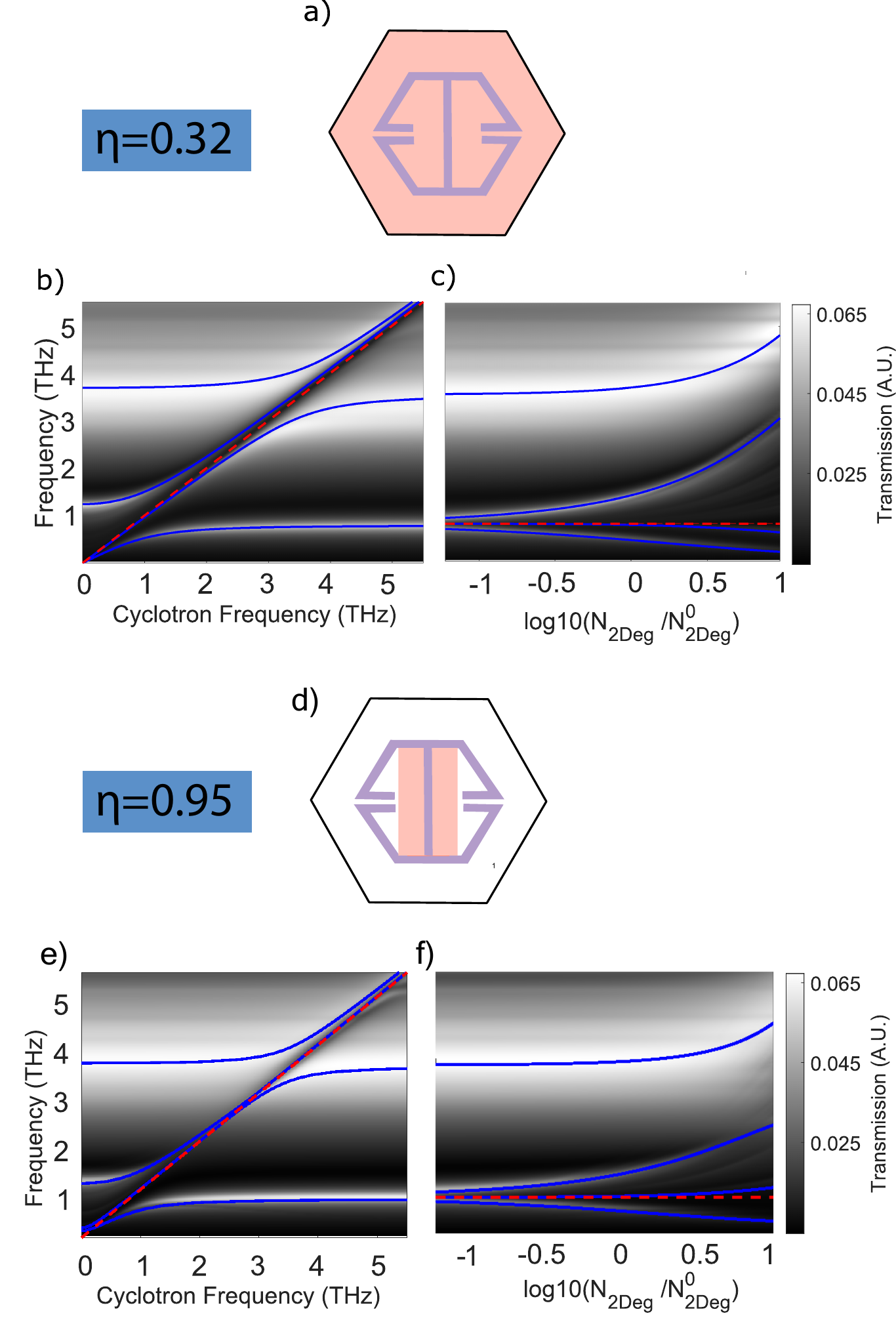}
\caption{\label{Fig3} Sketch of the structure including the hexagonal negative THz resonator (violet shape) fabricated on top of the GaAs substrate (white region), and the quantum well hosting the 2DEG  (light red layer), whose area occupies either the whole unit cell (a) or the central gap (d).
In the other panels we show the numerical calculations of the transmission as a function of the cyclotron frequency $\omega_c$ at a fixed electron density $N^0_{\textrm{2DEG}}=3 \times 10^{12}$ cm$^{-2}$ (b,f) and as a function of the electron density at a fixed cyclotron frequency $\omega_c=0.8$ THz (c,g). 
Panels (b,c) illustrate the results for the structure in panel (a), while panels (f,g) for the structure in panel (e). Insets in panels (b,f) show the calculated value of the overlap parameter $\eta_{2,1}$.
Blue solid lines highlight the fitted polaritonic resonances. 
}
\end{center}
\end{figure}
 We elucidate this insight by developing the first $M$ photonic functions $f_\nu(z,\textbf{r}_{\|})$ restricted over the quantum well plane as linear superpositions of at most $M$ orthonormal basis functions over the sample surface $\mathbb{S}$,
 \begin{eqnarray}
 \int_{\mathbb{S}}\phi^*_\nu(\textbf{r}_{\|})
 \phi_\mu(\textbf{r}_{\|})
 d\mathbf{r}_{\|}=\delta_{\nu,\mu}
  \end{eqnarray}
and obtain
 \begin{eqnarray} \label{phi}
 f_{\nu}(z,\textbf{r}_{\|})=\sum_{\mu\le \nu} \alpha_{\nu,\mu} \phi_\mu(\textbf{r}_{\|}).
 \end{eqnarray}
It is always possible to choose the basis such that $\alpha_{1,1}$ is real and $\alpha_{\nu,\mu}=0$ if $\nu < \mu$.
Using Eq. \ref{phi} the degree of non-orthonormality between the resonator modes with respect to the quantum well plane can be captured by defining the overlap matrix 
  \begin{equation}
  \label{k12}
   \mathcal{F}_{\nu,\mu}=\int_\mathbb{S}  f^*_{\nu}(z,\textbf{r}_{\|}) f_{\mu}(z,\textbf{r}_{\|})d\mathbf{r}_{\|}= \sum_{\gamma \le \min(\nu,\mu)} \alpha^*_{\nu, \gamma} \alpha_{\mu, \gamma},
  \end{equation} 
   and its normalized version
   \begin{equation} \label{eta}
   \eta_{\nu,\mu}=\frac{\mathcal{F}_{\nu, \mu}}{\sqrt{\mathcal{F}_{\mu,\mu} \mathcal{F}_{\nu, \nu}}},
   \end{equation}
both of which may assume values from $0$ to $1$. 
These matrices quantify the spatial overlap of any pair ($\mu,\nu$) of photon modes over the quantum well plane.
A diagonal matrix $\eta_{\nu,\mu}\propto \delta_{\nu,\mu}$ implies vanishing overlap between the photon modes, while a fully populated matrix corresponds to a strong overlap.

By introducing a set of collective bosonic matter operators
\begin{eqnarray}
\label{b1}
\hat{b}_{\mu}=\frac{1}{\sqrt{N_\text{2DEG}}}\sum_{j=1}^{N} \phi_{\mu}(\textbf{r}_{j,\|})\hat{c}_j, 
\end{eqnarray}
with the in-plane position $\mathbf{r}_{j,\|}$ of the $j$th electron, we can finally write the Hamiltonian in Coulomb Gauge as
    \begin{eqnarray}
    \label{finalH}
    \hat{H}&=&\sum_{\nu} \hbar \omega_{\nu} \hat{a}_{\nu}^\dagger \hat{a}_{\nu}+\sum_\nu \hbar \omega_c \hat{b}_\nu^\dagger \hat{b}_\nu\\
    &+&\sum_{\nu, \mu}  h_{\nu,\mu} \left ( \hat{a}_{\nu}^\dagger + \hat{a}_{\nu} \right)  \left ( \hat{a}_{\mu}^\dagger + \hat{a}_{\mu} \right)\nonumber \\
    &+&\sum_{\nu}   \sum_{\mu \le \nu}\left [\left(g_{\nu,\mu} b_{\mu} +g^*_{\nu,\mu} b_{\mu}^\dagger \right) \left ( \hat{a}_{\nu}^\dagger +\hat{a}_{\nu} \right)   \right  ].\nonumber
    \end{eqnarray}

Here, 
  \begin{eqnarray} \label{g}
  g_{\nu,\mu}&=& \alpha_{\nu,\mu}\sqrt{\frac{\hbar^2 \omega_c N_{\text{2DEG}} e^2 }{2 m^* \epsilon_0 \bar{\epsilon}_{r} \omega_\nu \mathbb{V}_{\nu}}}  ,\\
   h_{\nu,\mu}&=&\sum_{\gamma \le \nu,\mu} \frac{ g_{\nu,\gamma} g_{\mu,\gamma}}{\hbar\omega_c},\nonumber
   \end{eqnarray}
represent coupling parameters consisting of the background dielectric constant $\bar{\epsilon}_r$ of the quantum well material.
This Hamiltonian presents some important features. First, it is bosonic and quadratic, which allows us to determine its eigenmodes by Hopfield diagonalization \cite{hopfieldTheoryContributionExcitons1958}. Second, the interaction term displays cross-interactions between different photon and matter modes.

Following the Hopfield approach, we diagonalise the Hamiltonian by introducing the hybrid multi-mode polariton operators,
 \begin{eqnarray}
 \label{p}
 \hat{p}_{\mu}&=& \sum_\nu \left ( x_{\nu,\mu} \hat{a}_\nu +w_{\nu,\mu} \hat{b}_\nu +y_{\nu,\mu} \hat{a}_\nu^\dagger  + z_{\nu,\mu} \hat{b}_\nu^\dagger\right ), 
 \end{eqnarray} 
whereby  $\left(x_{\nu,\mu}, w_{\nu,\mu}, y_{\nu,\mu}, z_{\nu,\mu} \right)$ are real-valued Hopfield coefficients. 
The dressed polariton frequencies $ \omega^p_\mu$ are the eigenvalues of the polariton eigenequation
 \begin{eqnarray}
 \hbar\omega_{\mu}^p\hat{p}_\mu&=&\left[ \hat{p}_\mu,\hat{H}\right].
 \end{eqnarray} 
The Hopfield transformation can subsequently be inverted as
\begin{eqnarray}
\left (\hat{a}_\nu +\hat{a}_\nu ^\dagger  \right )&=& \sum_\mu \left (x_{\nu,\mu} - y_{\nu,\mu}\right)  \left( \hat{p}_{\mu} + \hat{p}_{\mu}^\dagger \right),
\end{eqnarray}
allowing us to find the coupled electric field components corresponding to the non-Hermitian vector potential 
 \begin{eqnarray} \label{coupled_fields}
  \hat{E}_{-}(\textbf{r})&=&\sum_{\nu, \mu}  \sqrt{\frac{\hbar \omega_\nu}{2 \epsilon_0 \bar{\epsilon}_{r} \mathbb{V}_\nu}}  f_{\nu} (\textbf{r})  \left (x_{\nu,\mu} - y_{\nu,\mu}\right)  \left(   \hat{p}_{\mu}^\dagger +  \hat{p}_{\mu} \right),\nonumber \\ 
  \hat{E}_{+}(\textbf{r})&=&\sum_{\nu, \mu}  \sqrt{\frac{\hbar \omega_\nu}{2 \epsilon_0 \bar{\epsilon}_r \mathbb{V}_\nu}}  f^*_{\nu} (\textbf{r})  \left (x_{\nu,\mu} - y_{\nu,\mu}\right) \left(   \hat{p}_{\mu}^\dagger +  \hat{p}_{\mu} \right).\nonumber \\
  \end{eqnarray}
From Eq.~\ref{coupled_fields} we can clearly see that, as expected from our initial discussion, the electric field corresponding to the polaritonic mode $\hat{p}_{\mu}$ is a linear combination of all bare electromagnetic mode profiles $f_{\nu} (\textbf{r})$, each weighted by the Hopfield coefficients. 

\section{Semi-Analytical Results}

In order to highlight the role of the normalised overlap factors for the coupling strength, we now assume a single pair of photonic modes ($M=2$) with  frequencies $\omega_1$ and $\omega_2$ and mode volumes $\mathbb{V}_{1}$ and $\mathbb{V}_{2}$. Their non-orthogonality is quantified by a single overlap parameter $\eta_{2,1}$. By expliciting Eq. \ref{k12}, we arrive at
\begin{eqnarray}
\mathcal{F}_{1,1}&=&\alpha_{1,1}^2,\nonumber\\
 \mathcal{F}_{2,2}&=&|\alpha_{2,1}|^2+|\alpha_{2,2}|^2, \\
 \mathcal{F}_{2,1}&=& \alpha^*_{2,1}\alpha_{1,1},\nonumber
\end{eqnarray}
which leads to 
\begin{eqnarray}
\alpha_{1,1}&=&\sqrt{\mathcal{F}_{1,1}} ,\nonumber \\
\alpha_{2,1}&=&\frac{\mathcal{F}_{2,1}^*}{\sqrt{\mathcal{F}_{1,1}}}=\sqrt{\mathcal{F}_{2,2}} \eta^*_{2,1}, \\
\alpha_{2,2}&=&\sqrt{\mathcal{F}_{2,2}-\frac{|\mathcal{F}_{2,1}|^2}{\mathcal{F}_{1,1}}}=\sqrt{\mathcal{F}_{2,2}} \sqrt{1-|\eta_{2,1}|^2}.\nonumber
\end{eqnarray}

Defining the renormalised mode volume as $\tilde{\mathbb{V}}_{\nu}=\frac{\mathbb{V}_{\nu}}{\mathcal{F}_{\nu,\nu}}$, Eq. \ref{g} leads to expressions for the coupling strengths
\begin{eqnarray} \label{g2}
  g_{1,1}&=& \sqrt{\frac{\hbar^2 \omega_c N_{\text{2DEG}} e^2 }{2 m^* \epsilon_0 \bar{\epsilon}_{r} \omega_1 \tilde{\mathbb{V}}_{1}}}  ,\\
  g_{2,1}&=& \sqrt{\frac{\hbar^2 \omega_c N_{\text{2DEG}} e^2 }{2 m^* \epsilon_0 \bar{\epsilon}_{r} \omega_2 \tilde{\mathbb{V}}_{2}}} \eta_{2,1},\\
  g_{2,2}&=& \sqrt{\frac{\hbar^2 \omega_c N_{\text{2DEG}} e^2 }{2 m^* \epsilon_0 \bar{\epsilon}_{r} \omega_2 \tilde{\mathbb{V}}_{2}}} \sqrt{1-|\eta_{2,1}|^2}.
\end{eqnarray} 
For the given basis, the interpretation of these coefficients is that the photonic mode $\nu=1$ is coupled to only a single matter mode, $\mu=1$. In contrast, the coupling strength for the photonic mode $\nu=2$ originates from simultaneous coupling to both matter modes owing to the non-vanishing overlap parameter $\eta_{2,1}$. 

In order to show the peculiar spectroscopic features expected in systems with non-negligible overlap between the photonic modes, we plot in Fig. \ref{Fig2} the spectra obtained by diagonalising the Hamiltonian in Eq.~\ref{finalH} for two resonator modes. 
The three cases concern settings of vanishing overlap ($\eta_{2,1}=0$, panels a,b), medium overlap ($\eta_{2,1}=0.5$, panels c,d), and maximum overlap ($\eta_{2,1}=1$, panels e,f), whereby in each case the left and right panel show spectra as a function of the cyclotron frequency, and electron density $N_{\textrm{2DEG}}$, respectively. 
 
We can point out two characteristic signatures for the overlap. First, we consider varying the cyclotron frequency (panels a,c,e).
For vanishing mode overlap $\eta_{2,1}=0$ (panel a), we observe the opening of separate polariton gaps for each pair of photonic mode and matter excitation. 
On the contrary, maximum overlap of $\eta_{2,1}=1$ (panel e) leads to the emergence of an \emph{S-shaped} resonance (blue curve).
In this case, the mode structure originates from the coupling of a single matter excitation $\mu=1$ to both photonic modes $\nu=1,2$, simultaneously, leading to three polariton branches in total. The \emph{S-shaped} center mode is confined between the cavity frequencies $\omega_1$ and $\omega_2$, thus manifesting a double-mode nature.
Second, we analyze the mode structure as a function of electron density (panels b,d,f).
Here, we see that at larger densities and thus larger couplings, two modes blue-shift in the case of vanishing overlap, while a single mode blue-shifts in the presence of substantial overlap.
We attribute this behavior to the contribution of the diamagnetic term which, being of higher order in $N_{\textrm{2DEG}}$, becomes dominant at very large densities and tends to blue-shift the upper polariton of each set of polaritonic solutions, taking into account that polaritonic modes never cross their bare components  \cite{deliberatoLightMatterDecouplingDeep2014,todorovDipolarQuantumElectrodynamics2015}.  Nevertheless, in the case of maximum overlap, the diamagnetic term between the two photonic modes leads to a repulsion of the upper polaritons, leading to an anti-crossing behaviour above a certain critical value of the electronic density.

\section{Numerical results}
 \begin{figure}[htbp!]
\begin{center}
\includegraphics[width=0.5\textwidth]{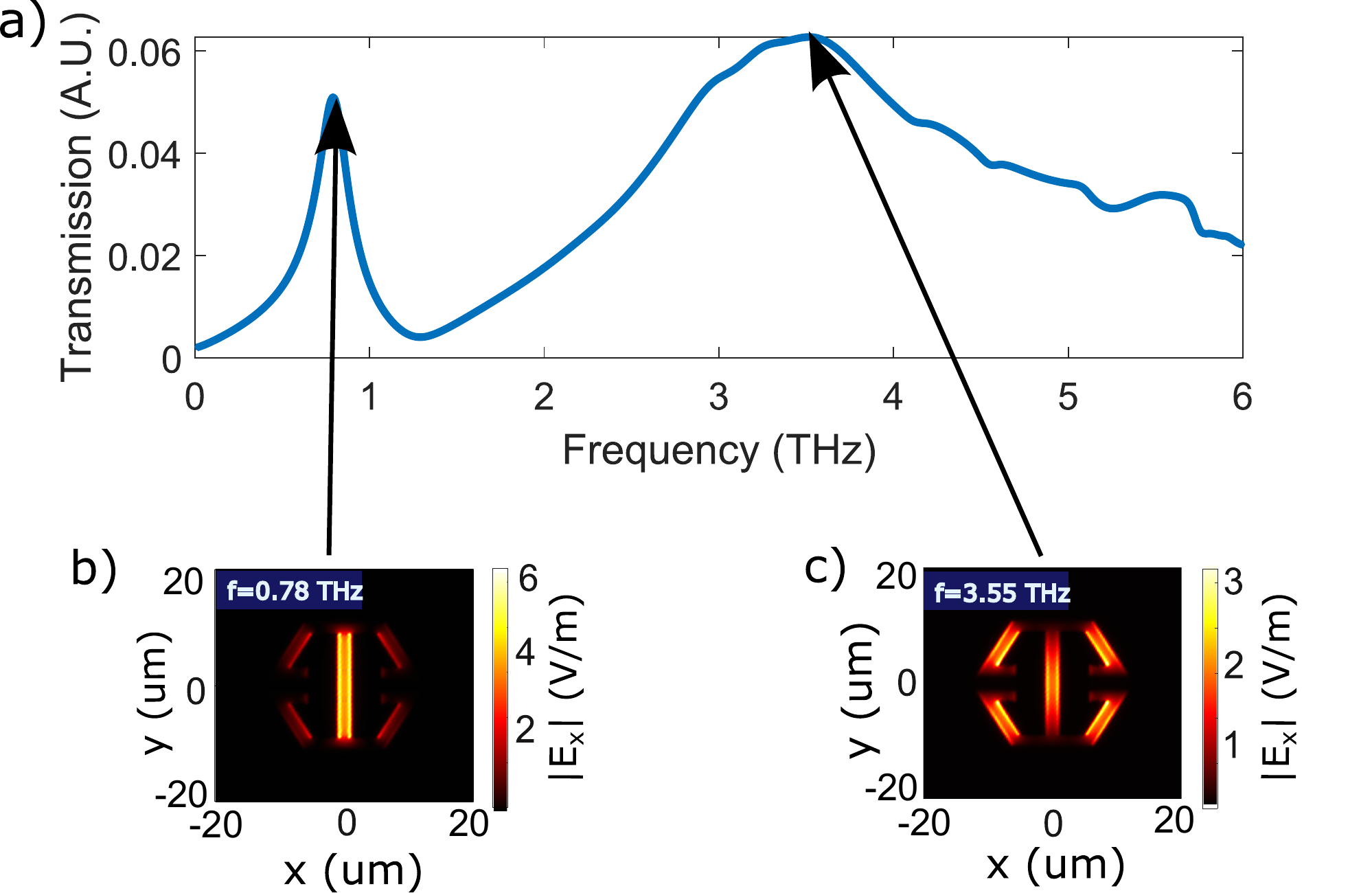}
\caption{\label{Fig4} 
 Transmission spectra for the resonator (a). The resonances with frequencies up to $5$THz are identified by black arrows and the corresponding in plane field distribution along the gap direction is plotted for each of the $M=2$ resonance in panel (a) with bare frequencies $\omega_1=0.78$ THz (b) and  $\omega_2=3.55$ THz (c). } 
\end{center}
\end{figure}

\begin{figure}[htbp]
\begin{center}
\includegraphics[width=0.5\textwidth]{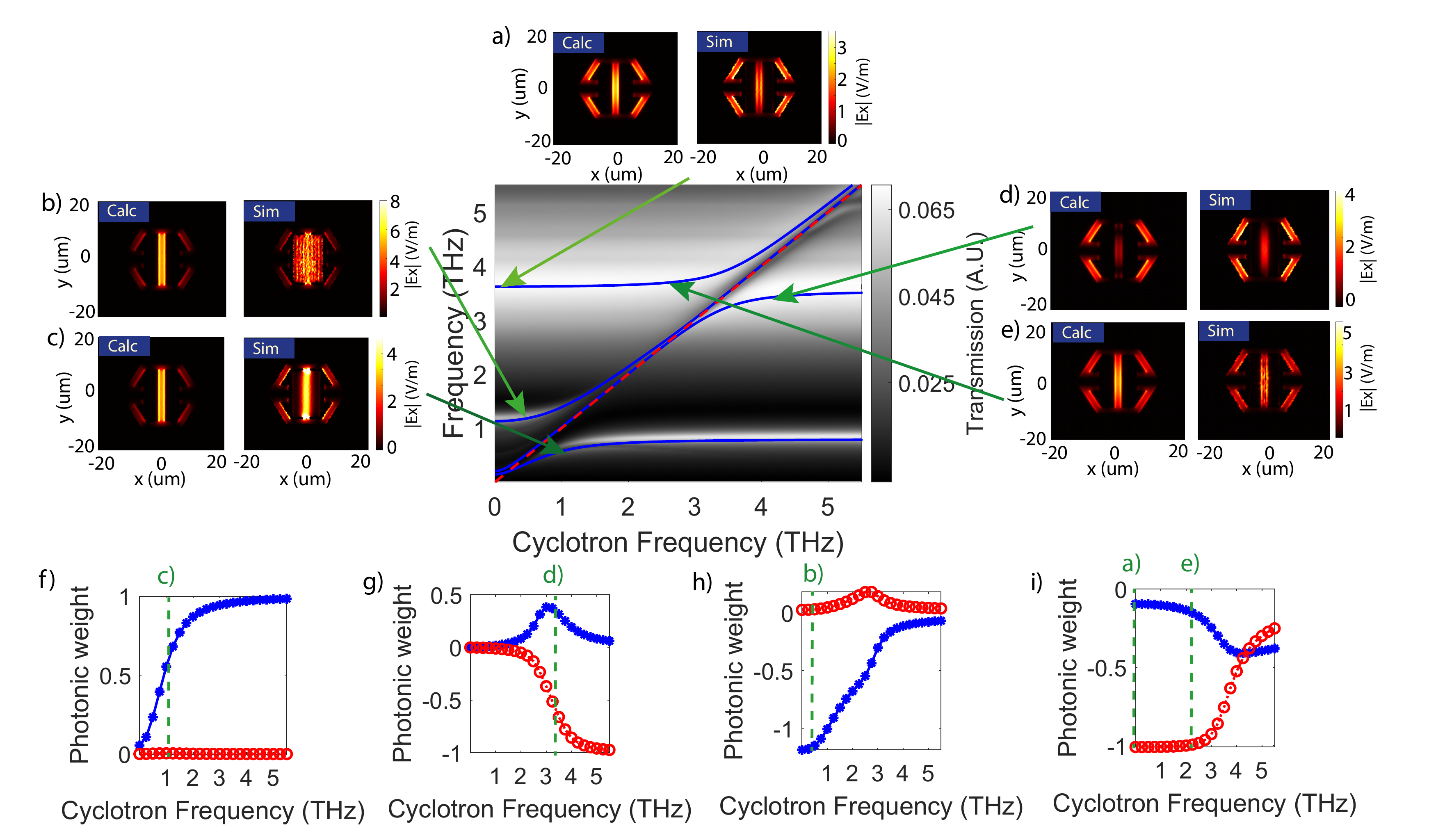}
\caption{\label{Fig5} Simulated Transmission spectrum as function of $\omega_c$ at $N_{\textrm{2DEG}}=3\times 10^{12}$ cm$^{-2}$ for the hexagonal resonator. The colormaps (a-e) represent the in-plane electric field profile along the gap direction $|E_x|$ (the y-component results negligible), extracted on the quantum well plane, corresponding to the coordinates marked by the green arrows.  The bottom panels (f-i) display the weight $|x_{\nu,\mu}-y_{\nu,\mu}|$ of the photonic mode $\nu$ in the polariton mode $\mu$, as appearing in Eq.~\ref{coupled_fields} for all the polariton modes, with the coordinates of the colormaps above marked by dashed green lines. The (f-i) plots are ordered following an ascendant order for the polariton frequencies.} 
\end{center}
\end{figure}

\begin{figure}[htbp]
\begin{center}
\includegraphics[width=0.5\textwidth]{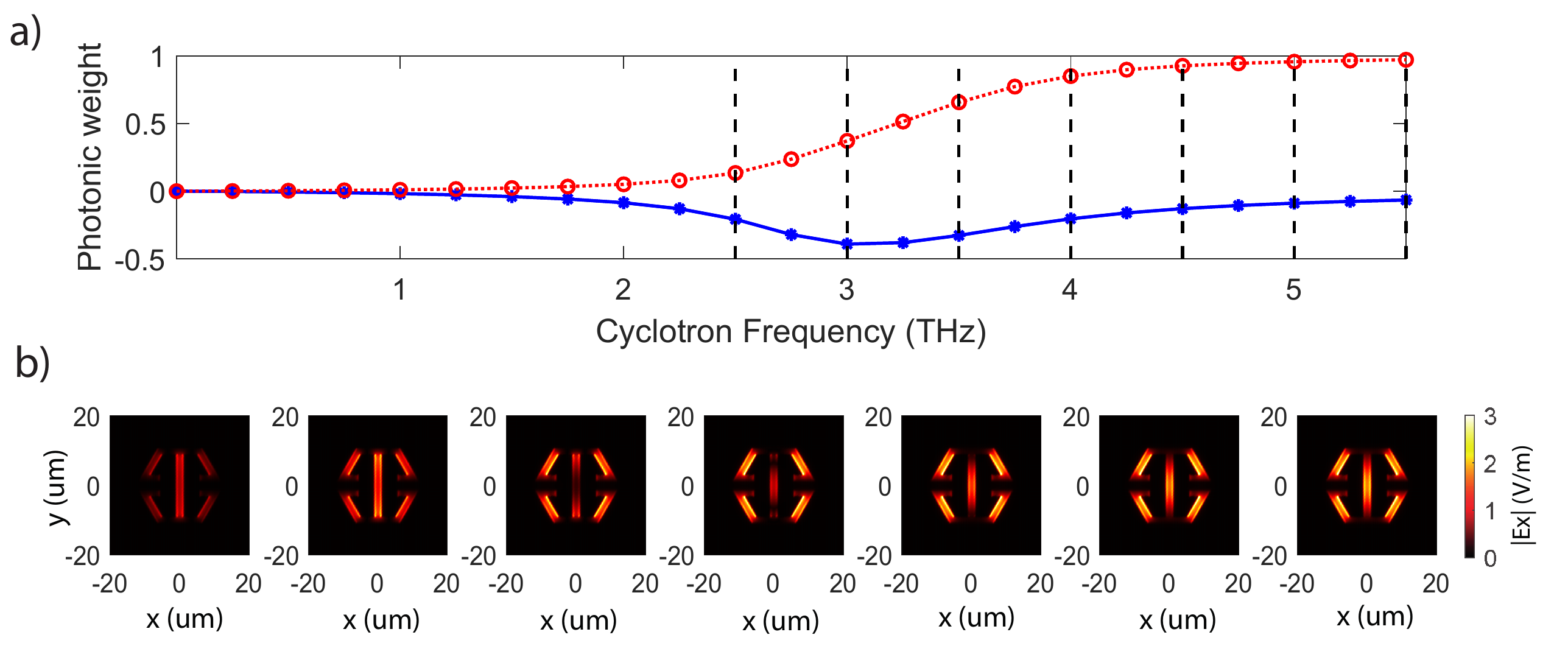}
\caption{\label{Fig6} Evolution of the electric field map for the second lower polariton mode. (a)  photonic weight components ($\nu=1$ blue dotted curve, $\nu=2$ red dotted curve) for the second coupled mode as a function of the cyclotron frequency at the fixed density $N_{\textrm{2DEG}}=3 \times 10^{12}$ cm$^{-2}$.
(b) in-plane electric field profiles corresponding to the cyclotron frequency coordinates marked by the vertical black dashed lines in (a). }
\end{center}
\end{figure}

In order to explore the relevance of our theory for experiments with Landau polaritons, we performed numerical FEM simulations of a metamaterial coupled to a 2DEG, following Bayer et al. \cite{bayerTerahertzLightMatter2017}.
To compute the complex field distribution and transmission spectra without any fitting parameters, our formalism models the dielectric environment of the nanostructure with a dielectric constant of $\epsilon_{\rm Au}=10^5+10^5i$ for the gold metamaterial \cite{bayerTerahertzLightMatter2017}.
The cyclotron resonance of the two-dimensional electron gas is implemented as a gyrotropic medium, where the dielectric tensor of a plasma of charge carriers magnetically biased along the z-direction describes the two-dimensional polarization response of the cyclotron resonance in the plane perpendicular to the magnetic field.
In the z direction we only employ the background dielectric constant, as the confinement inhibits a plasma response. Additionally, to reduce numerical complexity of modelling several quantum wells and corresponding barriers between them with finite thickness, we employ an effective medium for the complete quantum well stack with an effective dielectric tensor \cite{PhysRevB.80.245301}. In the x-y-direction, we employ periodic boundary conditions to reflect the array character of our structure and solve Maxwell's equations numerically. The resulting far-field calculations predict experimental results across the entire spectral range with high accuracy. 

Our structure is a \emph{negative} resonator (cut from a gold surface, Fig.~\ref{Fig3}) \cite{maissenUltrastrongCouplingField2014} of hexagonal shape, fabricated on top of a gallium arsenide (GaAs) substrate (white area in panels a,d), with the cyclotron resonances hosted in doped GaAs quantum well structures (light red region in panels a,d). 
In order to explore the direct impact of the overlap over the optical spectrum, we consider two types of quantum well designs. In the first layout the 2DEG covers the whole unit cell area (panel a). We refer to this design as \emph{unstructured}. A second layout is instead realised by in-plane confinement of the 2DEG within a small rectangular patch at the center of the resonator (panel d). We refer to this layout as \emph{structured}. 

The numerical transmissions for the four samples are shown in panels (b,e) as a function of the cyclotron frequency, and in panels (c,f) as a function of the electron density $N_{\textrm{2DEG}}$. 
 The simulation is performed considering an exciting electromagnetic wave which is linearly polarized along the gap (x) direction, and incident perpendicularly to the metamaterial plane.
 From the transmission spectrum at low electronic density, shown in Fig.~\ref{Fig4} (a), we recognize $M=2$ active photon resonances within the given frequency range, whose in-plane field profiles along the gap direction are plotted in Fig.~\ref{Fig4} (b,c).
 
In the structured case the patch acts as a Fabry-P\'erot resonator for the quasi-2D plasmonic excitations of the electron gas in the QWs
 \cite{PhysRevB.25.7603}.
 This leads to a non-vanishing frequency for the fundamental plasmonic mode to which the lower polariton in panel (e) of Fig.~\ref{Fig3} would converge for a vanishing cyclotron frequency.
We estimated the fundamental plasmon mode frequency using the formula \cite{popovResonantExcitationPlasma2005}
\begin{align}
\omega_P^0=  \sqrt{\frac{N_{\textrm{2DEG}} e^2 \pi}{2 m^* \epsilon_0 \epsilon_{\mathrm{eff}} W }},    
\end{align}
 with $W$ the patch width and $\epsilon_{\mathrm{eff}}$ the effective permittivity taking into account the screening of the gold resonator by averaging the screened and unscreened portions of the QW area. The resulting value is $\omega_P^0\approx 0.2\,$THz.
 Although for the sake of completeness we did use such a value in our simulations for the structured QWs, we notice that for such low frequencies the polaritons have vanishing photonic components and the transmission spectra are not noticeably affected by the exact value of $\omega_P^0$.

 Once we calculated the overlap parameter as in Eq. \ref{eta} for the two configurations, we employed our multimode theory to fit simultaneously the resonances for both the spectra of the $\omega_c$-sweep and the $N_{\textrm{2DEG}}$-sweep, considering the matter resonance as the magnetoplasmon mode $\tilde{\omega}_c$ and treating the normalised mode volumes $\tilde{\mathbb{V}}_{\nu}$ as fitting parameters. 
 
 From the discussion in the previous section we expect that passing from the unstructured to the structured sample, as the integration surface is reduced, not only the normalised mode volumes $\tilde{\mathbb{V}}_{\nu,\mu}$ will vary, but also the modes will become less orthogonal, thus increasing the overlap parameter $\eta_{2,1}$. This is indeed the case as can be seen by the calculated values of $\eta_{2,1}=0.32$ for the unstructured sample in Fig.~\ref{Fig3} (a) and of $\eta_{2,1}=0.95$ for the structured sample in Fig.~\ref{Fig3} (d) derived by Eq. \ref{eta}.

Comparing the transmission spectra for the different configurations allows us to recognize, albeit in attenuated form, the main differences in the spectral features predicted by the theory (marked on the plots by blue solid lines). At a first glance, we notice that the single polariton anticrossings are well resolved in the unstructured case, as they mainly arise from one-to-one coupling of photonic modes to orthogonal matter excitations. In the structured platform instead, we observe a reduction of the polariton splitting, and the appearance of an S-shaped resonance. The reduction of the polariton splitting is mainly due to the fact that reducing the integration area for the single mode leads to a larger normalised mode volume $\tilde{\mathbb{V}}_\nu$, and as such to a smaller coupling strength. On the other hand, the confinement of the 2DEG around the central gap of the resonator increases the overlap between the modes, which becomes close to 1, leading to the appearance of the characteristic S-shaped polariton.


We report in Figs. \ref{Fig5} the in-plane electric field distributions along the gap direction for the coupled modes of the hexagonal resonator platform in the structured configuration. 
The reported data sets are extracted from the FEM simulation and calculated by our multimode theory, respectively plotted at the right and left sides of panels (a-e).

 The simulation field maps are extracted at the resonance and at the value of $\omega_c$ marked by the green arrows on the central $\omega_c$-sweep transmission plot. The corresponding theoretical electric field profiles are instead obtained by Eq.~\ref{coupled_fields} as linear combinations of the numerically extracted fields of the uncoupled resonances, shown in Fig.~\ref{Fig4}, weighed by the photonic coefficients displayed by panels in \ref{Fig5} (f-i). Note that the linear superposition is calculated at a cyclotron frequency within half of the resonance linewidth from the nominal one. This procedure has been necessary because the value of the Hopfield coefficients, and thus the predicted shape of the electromagnetic mode, varies strongly within the linewidth of the broadened modes forming the upper anticrossing. 
 
 By observing the field maps we can notice that these refer to three different cases: panels (a,c,e) display map distributions similar to the uncoupled ones, as the weight of one of the two modes is greatly dominant over the other.
    Panel (d) displays a case in which the two photonic weights are comparable, and the electric field map is noticeably different from either of the bare ones. Finally, in panel (b) our theory predicts the field of the bare photonic mode mainly localised in the central gap, while the simulation shows the electric field diffracting in the far-field of the plasma waves, although remaining confined on the area of the quantum well patch. This effect is related to the one recently investigated in Ref. \cite{rajabali2021polaritonic}. Here, the authors point out how the electromagnetic field, confined in the resonator gap, can excite a continuum of propagative high-wavevector plasmonic waves leaking away energy from the polaritonic resonances. In our case the main difference is that the patch acts as a Fabry-P\'erot resonator. Even if higher-order discrete modes are quasi-resonant with one more polaritonic branches, the energy of the excited modes remains confined in the patch and has thus only a limited effect on the polaritonic resonances \cite{Shima}. 
    Our two-mode Hopfield model misses this effect, which could nevertheless be correctly described expanding the basis to include many discrete plasmonic modes of the patch \cite{Cortese:19} or alternatively using a theory able to deal with continuum spectra \cite{Cortese:2022}.
     
  Finally, Fig. \ref{Fig6} highlights the modification of the in-plane electric field driven by the multi-mode hybridization for the specific case of the second polariton mode, from the anticrossing point to its saturation at the higher photonic frequency $\omega_2=3.55$ THz. The calculated electric field maps in (b) refer to the cyclotron frequency values marked by the vertical black dashed lines in panel (a) (same as Fig. \ref{Fig5} (g)). We can clearly see how changing the cyclotron frequency varies the electric field map, displacing the minimum of the field across the central gap, a feature suggestive of potential applications in sub-wavelength sensing and optical tweezers. 
  
  Our results thus demonstrate that, by optimizing the resonator-2DEG structure, we are able to dynamically modify the sub-wavelength electromagnetic field profile, moving its maxima by varying the applied magnetic field.

\section{Conclusions}
In conclusion, we theoretically investigated the multi-mode coupling between the cyclotron resonances of a 2DEG and highly-confined THz-resonator modes. We developed a general theory describing multi-mode coupling taking into account the non-orthogonality of the electromagnetic modes. We highlighted specific spectral features due to the presence of multiple photonic modes and demonstrated the possibility to tune the level of inter-mode coupling by lateral confinement of the 2DEG. Finally using these effects opens up the possibility to dynamically tailor the spatial profile of sub-wavelength electromagnetic modes by varying the applied static magnetic field. This approach can potentially be used to realize sub-wavelength optical tweezers to trap and move nanoparticles over sub-micron distances.

The theoretical results encourage us to explore novel experimental methods and setups allowing to observe the predicted modification of the electric field profiles, driven by the coupling. 
Moreover, we aim to investigate further different resonator configurations, in order to maximise the effects of the multi-mode hybridization, heading towards novel quantum technological applications, based on a controllable and potentially dynamical tuning of high confined electromagnetic fields.

\section{Acknowledgements}
 S.D.L. is a Royal Society Research Fellow. S.D.L. and E.C. acknowledge funding by the Leverhulme Trust (Philip Leverhulme Prize) and the  Royal Society (grant RGF\textbackslash EA\textbackslash 181001). C. L. and J. M. acknowledge funding by the DFG through Grant No. LA 3307/1-2. 

\bibliography{bibliography.bib}

\end{document}